# Grid-Enabling Natural Language Engineering By Stealth

**Baden Hughes and Steven Bird**
Department of Computer Science and Software Engineering
University of Melbourne
Victoria, 3010, Australia
`{badenh, sb}@cs.mu.oz.au`

## Abstract

We describe a proposal for an extensible, component-based software architecture for natural language engineering applications. Our model leverages existing linguistic resource description and discovery mechanisms based on extended Dublin Core metadata. In addition, application design is flexible, allowing disparate components to be combined to suit the overall application functionality. An application specification language provides abstraction from the programming environment and allows ease of interface with computational grids via a broker.

## 1 Introduction

Computational grids are an emerging infrastructure framework for conducting research where problems are often data or processor intensive. A computational grid allows for large-scale analysis, distributed resources and processing, in addition to engendering new models for collaboration and application development. Foster et al, (2001, 2002) provides a physiological and an anatomical overview of grid computing services, and provides foundational architectures for application development in the grid space. Given the prevalence of large data sources in the natural language engineering domain and the need for raw computational power in the automated analysis of such data, the grid computing paradigm provides efficiencies otherwise unavailable to natural language engineering.

Language engineering applications are typically constructed out of several processing components, each responsible for a specialized task. Typical components include speech recognition, tagging, entity detection, anaphora resolution, parsing, etc. Each component is heavily parameterized and must be trained on very large datasets (e.g. the LDC Gigaword corpus (Graff, 2002)). Discovering optimal parameterizations is both data- and computationally-intensive. Building complex applications, such as spoken dialogue systems, depends on identifying and integrating suitable components often from a range of sources.

In this paper we describe a proposal for an extensible, component-based software architecture for natural language engineering applications which leverages linguistic resource discovery mechanisms and allows standard interfaces with computational grids for data analysis. An overview of the model is displayed below in Figure 1.

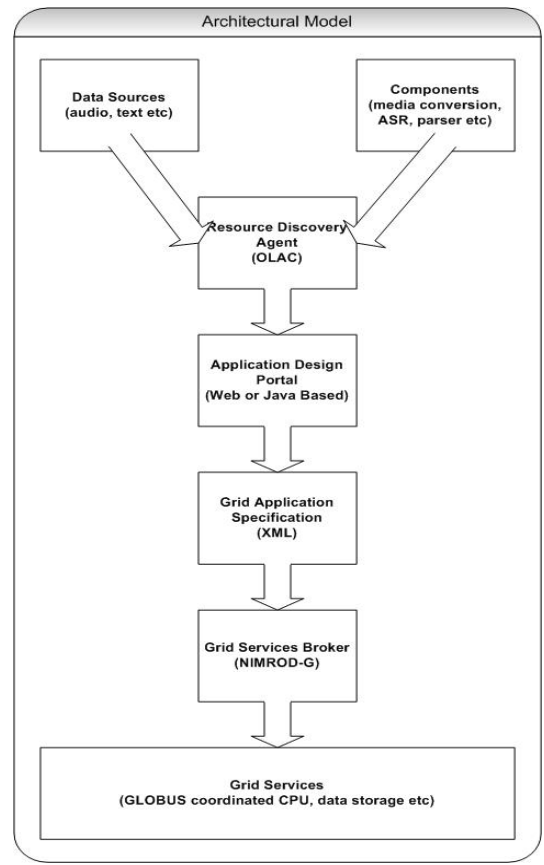

Figure 1. Architectural Model

Recent advances in the automated exposure and harvesting of data resource catalogues are proposed to be extended for the natural language engineering domain. Based on these extensions we can programmatically and intelligently discover data sources, components, applications and grid nodes offering specific services.

The model advocated here is broadly based on the concept of cloud computing (Siegele, 2001). In this model, a series of distributed individual components are assembled via an application framework, with internal communication requirements addressed by a common interface specification. An application consists of one or more data sources, together with a number of individual components, coordinated within an overall framework. Components are function specific implementations which adhere to a core series of standards for inter-component coordination which is open to extension by third parties.

Collections of components which form applications are then expressed using a grid application specification language which is parsed by a broker that interfaces with the computational grid infrastructure. Any application composed within or outside this model can address such a broker, and hence access the power and scalability of computational grids for processing.

This model has numerous benefits. First, existing natural language engineering applications can benefit from grid services without being grid-aware, simply by instructing the broker to perform a specified task as if it were a single server. Second, application definitions are a declarative specification of a processing task and of the relationship between processing components, facilitating substitution of equivalent components and aggregation of simple applications into more complex applications. Third, application definitions and result sets can be stored, described using standard metadata, and be discovered and re-used by other applications at a later date.

In this paper, we discuss the architectural foundations of our proposal, resource discovery mechanisms, component identification, multi-component application design (including several sample applications), the grid services interface and the grid application specification language. Finally some directions for future work are identified.

## 2 Basic Processing Paradigms

A range of software architectures for natural language engineering have been developed in the last decade, and matured to the extent that there are a wide range of applications built on foundations such as GATE (Cunningham et al, 1996, Cunningham 2002) and ATLAS (Bird et al, 2000). However, all of these software architectures envisage a local processing model, where data sources, entire applications, and computational resources are co-located. Additionally, processing is generally assumed to be serialized, with little opportunity for simultaneous processes taking advantage of analyses provided by each other.

Equally, there have been numerous developments for the deployment of computational grids, ranging from middleware, API's, libraries and infrastructural management approaches. Within the computational grid architectures currently deployed, there are both processor-centric and data-centric architectures, of which the most common is the processor-centric grid. In this context, grid aware applications typically adopt the approach of transporting the data from a storage site to the site of available CPU capacity. However in natural language engineering this is less attractive because the size of data sources is commonly large enough for any network based transport to be relatively inefficient and/or cost-prohibitive. The implications for the development of a broker architecture for data-centric grids are of orthogonal interest but are discussed briefly later.

The model advocated in this paper would work equally well in either context providing a sufficiently intelligent broker architecture was available. At the time of writing, it is envisaged that a processor-centric architecture will be utilized at least in the medium term.

## 3 Resource Discovery

Within any distributed natural language engineering system, there clearly exists a need to discover data sources, components, applications and processing services over the network. This need is expressed both by the application developer, who desires to build an application from a range of available components, as well as by the brokering agent, which needs to align application requirements with available computational resources.

There are a range of methods available to facilitate such discovery. Of note are the potential for the use of a Web Services implementation (using DAML-S or formal RDF), or embedding this task within the wider Semantic Web implementations (eg by W3C). Whilst the authors are appreciative of the developments within these areas, there already exists a language resource specific framework for resource description and discovery within the Open Language Archives Community (OLAC) which has significant benefit, and it is on this basis we proceed.

We propose to extend the OLAC resource description and discovery mechanisms. This will be achieved by adopting and extending standards based on the foundational work of Hughes (2002) for the encoding of information such as CPU and memory requirements, and component and application functionality. Electing to extend this standard allows us to leverage discovery tools and techniques already in existence and considered the benchmark for such processes. The OLAC Metadata Set (Bird and Simons, 2002a) is an extension of the Dublin Core Metadata Set (Dublin Core Metadata Initiative, 2003), thus ensuring widespread accessibility. The OLAC initiative has a standardized mechanism for extending OLAC metadata (Bird and Simons, 2002b) which we will adopt to describe data sources, applications and processing nodes. Having based resource descriptions on OLAC standards we can then query an OLAC aggregator to discover the existence and status of resources of interest.

## 4 Component Identification

In order to explore the grid computing paradigm through a natural language engineering framework, a component based model has been adopted. Within this model, we identify and describe a number of re-usable component types which can be combined to create multi-component applications to be executed in the grid environment. Each of these component types has as a common functional core the ability to communicate with a central management infrastructure using standard messaging interfaces. In developing prototype implementations, we intend to wrap existing components wherever possible.

### 4.1 Annotation Server

A large class of language engineering tasks involving time-series data can be construed as adding a new layer of annotation to existing data. For example, a process which takes speech input and produces text output (e.g. a speech recognizer) can be viewed as adding textual annotation to audio data, with the result that both data types remain accessible for further analysis. When several tasks operate in this mode, they collectively build a rich store of linguistic information. For example, a prosody recognition component could identify major phrase boundaries in spoken input, and a parser could employ both the ASR and prosody output in constructing parse trees.

Annotation graphs can be used to represent a diverse range of time-series annotations, including ASR output, POS tags, named entities, syntactic chunks or trees, aligned translations, dialogue acts, and so on. Importantly, the intermediate stages of many processing tasks are well-formed as annotation graphs, so certain outputs may become available shortly after the process starts, facilitating efficient pipelining and streaming. The Annotation Graph Toolkit (AGTK) (Linguistic Data Consortium, 2001-2003) is an open source implementation of annotation graphs which works with a mature API and a wide range of supported models and formats. The Annotation Server is a component that collects, collates and stores annotation graphs which may be generated or accessed by other components and applications and extends on proposals previously described by Cieri and Bird (2001), and available as a prototype in the current AGTK distribution.

### 4.2 Alignment

An Alignment component is used to forcibly align digitized speech with a supplied transcript. This can be performed at various levels of granularity, and partial results can be naturally represented as an annotation graph.

### 4.3 Automatic Speech Recognition

An ASR component can be used to create a time-aligned transcript, again represented directly as an annotation graph.

### 4.4 Data Source Packaging

A Data Source Packaging component divides data sources into logical units which can be distributed across the grid as individual processing tasks. As an example, consider a digital audio file of 500Mb in size. As this would be non-trivial to transfer to a remote processing node, a Packager could divide this data source into fifty 10Mb digital audio files, each of which could be compressed, transported and processed separately.

### 4.5 Conversion

A Conversion component is used to convert between media formats, character encoding schemes, and annotation types. The expected input and output formats of existing natural language engineering components are often incompatible, and this component will facilitate component integration, simplifying the task of wrapping existing components. In the case of media formats, this component will be able to serve a static data source as a stream source and vice versa. In the case of annotation conversion, this component will be able to convert between the wide variety of existing time-series annotation formats understood by AGTK.

### 4.6 Text Annotation

A Text Annotation component augments existing annotated text with a new layer of annotation, e.g. POS tags, sense tags, named entities, syntactic chunks and so forth.

### 4.7 Lexicon Server

A Lexicon Server component collects, collocates and stores lexical data that is generated or accessed by other components. Examples include the TIMIT lexicon (words and pronunciations from the TIMIT database) and WordNet. In the absence of a universal lexicon API there will need to be a family of commonly supported APIs.

### 4.8 Semantic Mapping

A Semantic Mapping component constructs and collates content based on theme identification. Themes may be defined by a list of keywords of concordance based output.

## 5 Multi-Component Application Design

Having wrapped such components, we propose to combine them to create a range of multi-component applications which can be executed over the distributed grid infrastructure. It is envisaged that applications will be assembled using a portal based approach for ease of deployment. Based on the components described earlier, we can build a number of different applications and application types. Next we discuss two of these applications, namely spoken passage retrieval and collaborative annotation.

### 5.1 Spoken Passage Retrieval

The Spoken Passage Retrieval application will allow a user to identify passages of interest within spoken document collections. This depends on the source having been transcribed and indexed. Transcription involves media conversion and ASR on the spoken document collection, with results stored in the annotation server. If transcripts are already available, then the ASR process is replaced with alignment. Next, the annotations must be indexed, preferably with the aid of a lexical resource such as WordNet. The resulting index will be viewed as a kind of lexicon, which permits lookup on words to give document regions. The above two processes can be executed in parallel on a grid architecture. The final application is not grid-based, but simply accesses a new database created by the grid (the spoken document index).

### 5.2 Collaborative Annotation

There are four basic types of collaborative annotation, and we plan to support them all. First, peers may cooperate in the construction of a large annotated resource, dividing up the workload and each performing the same task on their parcel of work. Limited coordination is necessary in order to ensure that all items are annotated appropriately (e.g. that 10% of items are doubly annotated for reliability testing). In the second type of collaboration, a supervisor vets the work of hired annotators, and optionally adds further codes that depend on their own critical analysis of the data. Third, two specialists may annotate the same dataset according to different theoretical models, in order to investigate the empirical differences between the theories and the extent to which categories posited by one theory can be derived from the other. Finally, collaboration may take place between human and automatic agents, which monitor human decisions and, over time, develop increasingly refined models that predict those decisions and expedite the work. The latter case is of significant interest in the grid context where automated agents may rapidly annotate large datasets over the computational grid, and hence be able to hypothesise language models in much shorter periods of time, which in turn can be used by human annotators.

## 6 Interfaces to Grid Services

The core infrastructure requisite for grid services can be provided in a number of ways. Historically, bootstrapping approaches such as those exhibited by PVM (Geist et al, 1994) and MPI (Message Passing Interface Forum, 2001) have gained widespread adoption. For our purposes, owing to their low level interaction with applications, the requirement that significant development is required to integrate these, and the technical management overhead of network resources that is required to gain full use of such frameworks, we have decided that there are reasonable alternatives.

Instead, we have adopted the approach provided by Globus (Globus Project, 2003). The Globus model allows for a distributed set of resources (the grid), upon which applications can execute in parallel or parametrically. Furthermore, within the Globus model, there is capacity for lightweight, cross-platform implementations which significantly appeal to the natural language engineering context where applications are often tied to a particular platform or architecture. Coordination of grid services is an included component within the Globus framework, through the notion of the virtual organization (VO) and accessible directly, through a middleware layer or by broker services which manage processes within the VO.

The model of Grid service interaction we have selected is to use a broker architecture which completely manages all aspects of grid interaction. Although this can be construed as "grid enablement by stealth", using a broker allows effort to be expended within the domain of interest (in this case natural language engineering) without concern for underlying grid infrastructure issues.

In our proposal, we adopt the broker model for managing all grid interface tasks. This simplifies application development as well as leveraging existing research in the area of broker architectures for grid environments. Within the grid environment, there is an emerging consensus that the NIMROD-G (Buyya et al, 2001) broker architecture will become the de-facto standard. Significant research, implementation and testing has been carried out using NIMROD-G through the Globus project and through World Wide Grid testbed (World Wide Grid, 2003). As such, it represents a hardened, real-world broker architecture on which we elect to build, rather than building our own broker, an area somewhat orthogonal to our primary research interests. Future collaborative work may extend into NLE-specific broker design (see Section 8 for some exploratory ideas).

The NIMROD-G architecture combines broker functions with grid services, for example, automated discovery of resources, as well as negotiation for services and dispatch management. In the context of this project, we will extend the NIMROD-G Resource Broker Architecture to allow for a standards-based interface by which the broker can parse the application specification language described earlier. This interface will be independent of application framework and hence available to any NLE framework which required grid based services for processing tasks.

As mentioned earlier, current grid processing models are processor-centric, yet natural language engineering data sources lend themselves more favorably towards data-centric implementations. Current implementations of NIMROD-G are also constrained by this assumption, and as such will need to be extended to allow data-centric processing. However, the extension of an existing broker architecture is significantly more appealing than writing a customized natural language engineering grid service broker from scratch.

## 7 Grid Application Specification

Our proposed Grid Application Specification Language is based on the model specified in Buyya (2003), but will be implemented in pure XML. In our model, the NIMROD-G broker architecture will accept parameterized input based on the natural language engineering application requirements, and will create the independent processing tasks, schedule them for processing on the grid, facilitate inter-task coordination and collate the results. Our

Grid Application Specification Language will allow interaction with a NIMROD-G application prototype and includes the ability to substitute variables both statically (at commencement) and dynamically (as the result of other tasks).

The Grid Application Specification Language will support the description of data requirements, processing requirements and communication requirements, both for individual components and aggregates. Data requirements include the location of static corpora and the interfaces for "dynamic" corpora (those accessible via a public API), and any requirements for media format, encoding scheme or annotation type which can be selected from or derived from the corpora. Processing requirements include the CPU, memory, storage and any task completion deadlines. Communication requirements include estimated bandwidth for the channels connecting the components. Additionally, information about the canonical ordering and dependencies between tasks that interact within the application will be documented here.

In addition to this basic functionality, we envisage three kinds of extended functionality: aggregation, dependency resolution, and storage.

Components can themselves be combined into pseudo applications with an overarching Grid Application Specification, such as may be the case where media conversion and encoding conversion are both required prior to other components becoming useful in the application context.

We have also identified the need for the broker to evaluate applications submitted to it for unresolved dependencies or incompatibilities. For example, the broker should be enabled to determine that the output of component A will be incompatible with input formats acceptable to component B, and intervene with a proposed solution (eg encoding conversion).

Furthermore, the Grid Application Specifications will be stored for later discovery. In order to meet this requirement, the Grid Application Specification must include sufficient information so as to comply with the discovery mechanism, namely metadata based on that standardized by the Open Language Archives Community. Through the use of linguistic metadata, including language technology extensions, we can fully describe the functionality of a resource such as on specified using the Grid Application Specification language. Automated discovery will occur through the use of a static repository harvester based on OLAC's Viser (Simons, 2002).

## 8 Future Directions

The architecture design allows for a number of immediately identifiable extensions. The first is the potential for a range of software architectures in NLE to generate application definitions, and as such interface with grid services. The second is a portal implementation that allows the user to select the data sources and types of analysis which are of interest and perform these transparently over the grid using the broker as an intermediary. The third is that grid resource broker architectures can be extended to encompass data-centric processing which will significantly impact on the application of grid services to natural language engineering. The fourth is the implicit vision that a number of entities will offer grid enabled natural language engineering services which participate in a virtual organization and can be discovered automatically and hence allow for specific analysis types.

In addition to this infrastructure, new components and applications within this framework will be developed. As an example, we can identify a 'Comparative Evaluation' application which would use Grid services to allow the user to build a large number of language models based on controlled, parameterized options and compare these.

Yet another area for future cross-disciplinary research is in the design of grid processing models themselves. Skillicorn (2001) identifies a number of computational grid types, of which the most common is the processor-centric grid as discussed earlier. In natural language engineering this model is less attractive because the size of data sources is typically significant enough for any network based transport to be relatively inefficient and cost-prohibitive. Skillicorn advocates the concept of the data-centric computational grid, where grid aware applications are transported to the location of the data source and describes a number of implications for architectural models relevant to natural language engineering. In this area, the requirements of natural language engineering may provide the catalyst for such an innovative broker design.

## 9 Conclusion

Computational grids enable the sharing and aggregation of geographically distributed resources for solving large-scale, resource and data intensive problems such as those found in natural language engineering. In this paper we have outlined a proposal for a software architecture which allows the ease of integration of such services with traditional natural language engineering development environments. The grid-enabled component-based architecture described in this paper is novel for a number of reasons.

Firstly, although the component based model is not necessarily innovative, the underlying motivation for this model (namely, ease of interface with grid services) brings a new perspective to the model design at both a component and application levels. Not only does it allow larger, more complex and computationally intensive processes to be divided for more efficient processing, but it also allows extrapolation for a slim-line application design which may be utilized in non-grid enabled contexts. In turn, this may allow efficiencies to be gained where previously computational power, data storage or bandwidth imposed constraints.

Secondly, the use of an application specification language is proposed, which allows our component based model to form parameterized input for a grid service broker, whilst allowing a wide range of underlying frameworks to access similar services. Applications can now potentially be evaluated for optimality and interdependency prior to execution, thus allowing for processing resources to be utilized more efficiently. Previously collated applications can be used as components in later, larger applications.

Thirdly we identify the need for existing grid service broker architectures to allow for data-centric processing, as opposed to currently implemented processor-centric approaches. The adoption of a broker based architecture and the open, accessible nature of the broker interface will allow a large class of natural language engineering applications to utilize grid services with corresponding gains in computational efficiency.

Finally the optional availability of previously compiled component sets, application definitions and output results are all made available as discoverable resources for future application design.

## Credits

The research described in this paper has been partially funded by the National Science Foundation under Grant Nos. 9910603 (International Standards in Language Engineering), 9978056, 9980009 (Talkbank) and by the Victorian Partnership in Advanced Computing Expertise Grant EPPNME092.2003.

The authors would like to thank Rajkumar Buyya for comments on earlier drafts of this paper.